\documentclass[amssymb,prb,twocolumn,superscriptaddress,floats,showkeys,showpacs]{revtex4-2}
\usepackage[T1]{fontenc}
\usepackage{authoraftertitle}
\usepackage{bm}
\usepackage{graphicx}
\usepackage{amssymb}
\usepackage{leftidx}
\usepackage{upgreek}
\usepackage{siunitx}
\usepackage{amsfonts}
\usepackage{amsmath} 
\usepackage{hyperref} 
\usepackage{multirow}
\usepackage{makecell}
\usepackage{xspace}
\usepackage{comment}
\usepackage{xr}
\usepackage{tabularx}
\usepackage{array}
\usepackage{adjustbox}
\usepackage{newtxmath}
\usepackage{soul}
\usepackage{textcomp}
\hypersetup{colorlinks,citecolor=blue, filecolor=blue ,linkcolor=blue , urlcolor=blue, pdftex}
\usepackage{lineno}
\usepackage[table]{xcolor}
\usepackage{ulem}

\def \crsbr{CrSBr}

\begin{document}

\title{Magneto-Excitonic Duality From Monolayer to Trilayer \crsbr}


\def \FUW{University of Warsaw, Faculty of Physics, Pasteura 5, 02-093 Warsaw, Poland}
\def \UCT{Department of Inorganic Chemistry, University of Chemistry and Technology, Technická 5, 166 28 Prague 6, Prague, Czech Republic}

\author{Igor Antoniazzi} 
\email{igor.antoniazzi@fuw.edu.pl}
\affiliation{\FUW}
\author{\L{}ucja Kipczak}
\affiliation{\FUW}
\author{Bruno Camargo}
\affiliation{\FUW}
\author{Gayatri}
\affiliation{\FUW}
\author{Chinmay~Mohanty}
\affiliation{\FUW}
\author{Kseniia~Mosina}
\affiliation{\UCT}
\author{Zden\v{e}k Sofer}
\affiliation{\UCT}
\author{Adam Babi\'nski}
\affiliation{\FUW}
\author{Arka Karmakar}
\email{arka.karmakar@fuw.edu.pl}
\affiliation{\FUW}
\author{Maciej R. Molas}
\email{maciej.molas@fuw.edu.pl}
\affiliation{\FUW}

\begin{abstract} 

Two-dimensional (2D) layered magnetic materials (LMMs) are a newly emerging class of van der Waals materials, opening new opportunities to study magneto-excitonic  coupling.
The air-stable, structurally and optically anisotropic A-type antiferromagnetic chromium sulfur bromide (\crsbr) is one of the most prominent examples of such LMMs.
We investigate photoluminescence (PL) and PL excitation of mono- to tri-layers \crsbr~and find that it exhibits a unique duplexity, supporting both Frenkel- and Wannier-Mott-like excitons.
Our magneto-optical experiments reveal a similar excitonic response from the mono- and trilayer systems and a completely different signature in the bilayer flake.
This shows a different origin of the low-lying excitonic species (A, A\textquotesingle~and B) in the band structure.
We confirm the robustness of the magneto-excitonic coupling in few-layer \crsbr.
Our work enables a more comprehensive exploration of the dual excitonic behavior in 2D materials.

\end{abstract}



\maketitle


\subsection*{Introduction \label{sec:Intro}}

The two-dimensional (2D) van der Waals (vdW) material family spans of insulators, semiconductors, (semi)metals, and superconductors~\cite{Novoselov2004, Geim2007, Geim2013, Manzeli2017, Mak2019}.
In 2017, a new member, intrinsic layered magnetic materials (LMMs), was experimentally confirmed as a vdW system~\cite{Gong2017, bevin2017}.
Since then, a race has been going on to understand and control its magnetic properties for next-generation ultrathin optoelectronic device applications~\cite{xiaomu2015, gang2018, Blancon2018, Shuqing2021, Pawbake2023nano, Komar2024, Smiertka2025}.
However, most of the newly discovered LMMs are highly chemically unstable under atmospheric conditions~\cite{Gong2017, telford2020, Torres2023}, making them unsuitable for real device applications.

\begin{figure*}[ht]
    \centering
    \includegraphics[width=.9\linewidth]{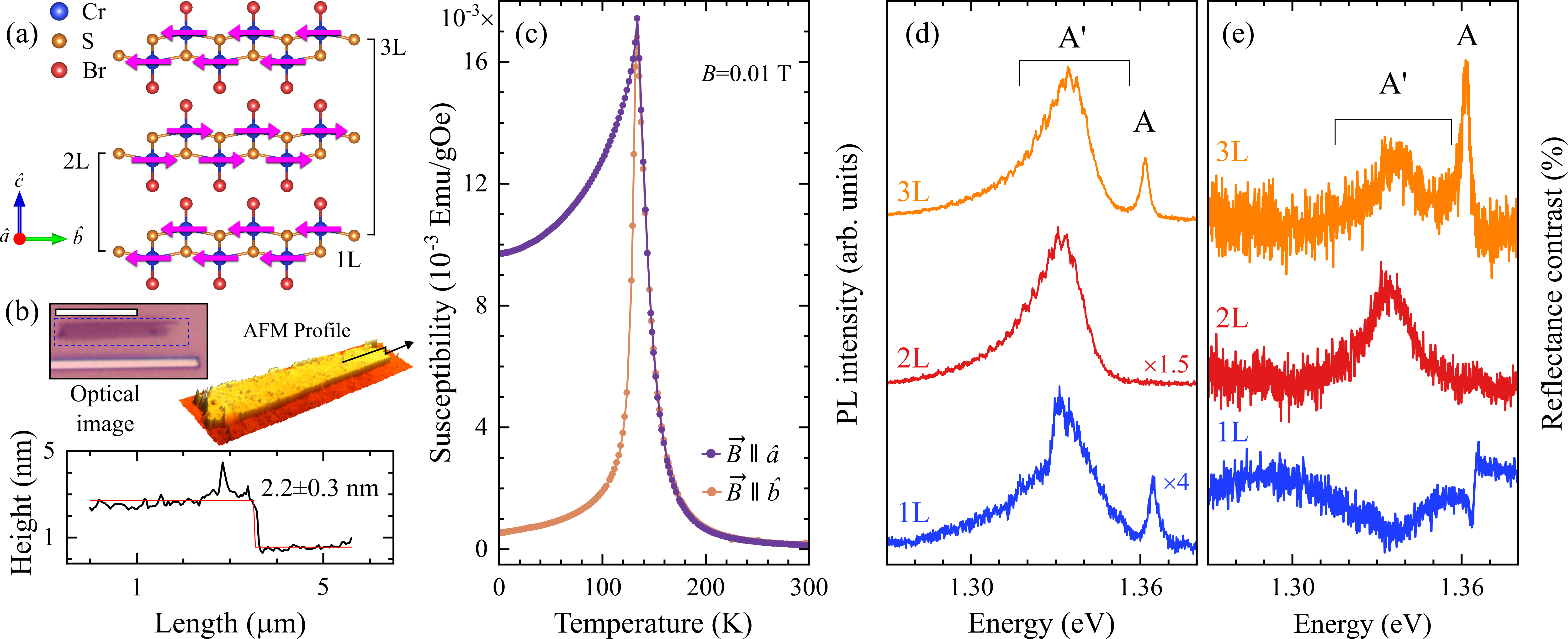}
    \caption
    {\label{fig1}
    (a) Illustration of the atomic structure of 1L, 2L, and 3L \crsbr.
    The $\hat{a}$, $\hat{b}$, and $\hat{c}$ denote the crystallographic axes.
    Pink arrows indicate the magnetic moments located on Cr atoms, giving rise to magnetic ordering.
    (b) Optical and atomic force microscopy of the 1L \crsbr~accompanied with its measured height profile, along the black arrow, shown at the bottom.
    The 1L flake is denoted by the blue dashed rectangle.
    The white bar represents a 5 $\upmu$m scale bar.
    The red line stands for the step function fitting. 
    (c) Temperature-dependent magnetic susceptibility ($\upchi$) for a \crsbr~single crystal in two orientations of the applied magnetic field ($\Vec{B}\parallel\hat{a}$ and $\Vec{B}\parallel\hat{b}$) measured in the FCC protocol under $B_\textrm{ex}$=0.01~T. 
    (d) Photoluminescence and (e) Reflectance contrast spectra of the 1L, 2L, and 3L CrSBr detected at $T$=5~K.
    The PL spectra were measured under excitation of 2.41~eV and with an average power 100 $\upmu$W.
    Data is vertically shifted and scaled by the indicated factors.
    }
\end{figure*}

Air-stable~\cite{Torres2023} chromium sulfur bromide (\crsbr)~is a vdW material crystallized in a low-symmetry orthorhombic crystal structure with strong uniaxial character, as evidenced in its structural and transport properties~\cite{goser1990, telford2020, sara2022, Wu2022}. 
It became the favorite choice to study the coupling between magnetic and optical properties~\cite{Torres2023, Pawbake2023, Pawbake2023nano}.
\crsbr~is an LMM compound with spin aligned in the layer plane along the $\hat{b}$ axis (magnetic easy axis) resulting in the ferromagnetic (FM) arrangement~\cite{goser1990, Katscher1966, beck19xx, francisco2023}.
The adjacent \crsbr~layers are coupled antiferromagnetically giving rise to \mbox{A-type} antiferromagnetism (A-AFM; the spin directions in the neighboring layers are opposite).
As shown in Fig.~\ref{fig1}(a), the \crsbr~monolayer (1L) presents a FM order, its bilayer (2L) is a perfect AFM system, while the trilayer (3L) and thicker material (>3L) are complex AFM structures.
Unlike other LMMs, the Néel temperature ($T_N$) of \crsbr~increases from about 132~K in bulk to about 140~K in 2L~\cite{goser1990, kihong2021}.
The highly anisotropic structure of \crsbr~is further reflected in its optical properties, which exhibit specific optical selection rules.
The electronic transitions are linearly polarized along the $\hat{b}$ axis~\cite{nathan2021, Klein2022}.
Similarly to other semiconducting vdW materials~\cite{Molas2019}, the optical response of \crsbr~is also dominated by excitons, but its nature is more complex due to its magnetic properties~\cite{nathan2021, Cenker2022, Shao2025, Christin2025, Smiertka2025}.
Although most of the semiconducting layered materials have delocalized Wannier-Mott excitons~\cite{Wannier1937, Molas2019}, the magnetic insulators support localized Frenkel excitons~\cite{Dirnberger2022}.
Bulk \crsbr~stands out by exhibiting an unusual duality in its excitonic dynamics~\cite{Smiertka2025}, supporting both Frenkel- and Wannier-Mott-like excitons.
Consequently, the rich magneto-excitonic coupling observed in bulk \crsbr~includes moderate redshifts of the Frenkel-like excitons, following a typical quadratic magnetic field energy dependence, along with giant magnetic shifts associated with Wannier-Mott-like excitons~\cite{Komar2024, Smiertka2025}.

In this study, we explore the magneto-excitonic properties of the \crsbr~with thicknesses ranging from 1L to 3L.
We investigate the layer-dependent photoluminescence emission (PL) and PL excitation (PLE) at low temperature ($T$=10~K) under magnetic fields ($B_\textrm{ex}$).
Under an external magnetic field (applied in $c$-axis) the excitonic features redshift up to the critical magnetic field of 2~T ($B_c$). 
The $B_c$ value is a consequence of the transition from the in-plane AFM ordering at 0~T (for 1L, the in-plane FM) to the out-of-plane FM phase.
The 1L/3L systems show two prominent PL peaks: a sharp A peak and a broad A\textquotesingle.
The former (A), has been rarely observed in the 1L system~\cite{Klein2023, francisco2023, tabataba2024, Shao2025}.
Whereas, 2L shows only a broad A\textquotesingle~peak.
Nevertheless, \crsbr~challenges the conventional separation of Frenkel and Wannier excitons, opening new opportunities for the study of hybrid excitonic behavior in 2D materials.

\subsection*{CrSBr thickness identification}

\crsbr~has a rectangular structure with unitary vectors $\hat{a}$=0.35~nm, $\hat{b}$=0.48~nm, and $\hat{c}$=0.79~nm, which creates an orthorhombic symmetry $Pmmn$ (D$_\textrm{2h}^\textrm{13}$) and a point symmetry C$_\textrm{2v}$~\cite{goser1990}.
Optical and atomic force microscopy images of the exfoliated 1L \crsbr~are shown in Fig.~\ref{fig1}(b).
The detailed description of \crsbr~growth and the preparation of its thin layers is given in the Methods section.
The measured height profile, presented in the lower panel of Fig.~\ref{fig1}(b), reveals a thickness of 1L.
In order to avoid any structural damage/doping, we did not perform any annealing/chemical treatment to get rid of the polymer residue. 
The measured thickness (2.2$\pm$0.3~nm) has an extra height of about 1.4~nm, which we ascribed to leftovers from the exfoliation process and the trapped water layer, as previously observed in other 2D materials~\cite{Blake2007, zhihong2007, Casiraghi2009, bertolazzi2013, junjie2015}.
We confirm the 1L nature by optical contrast and emission spectrum, as described in the following sections.
Similarly to 1L, the 2L and 3L flakes also showed extra height (see the Supplementary Information (SI)).
The precision of our thickness assignment is based on the characteristic optical properties of 2L \crsbr~\cite{francisco2023, Shao2025}.

\subsection*{Magnetic susceptibility of the \crsbr} 

\begin{figure*}[ht]
    \centering
    \includegraphics[width=1\linewidth]{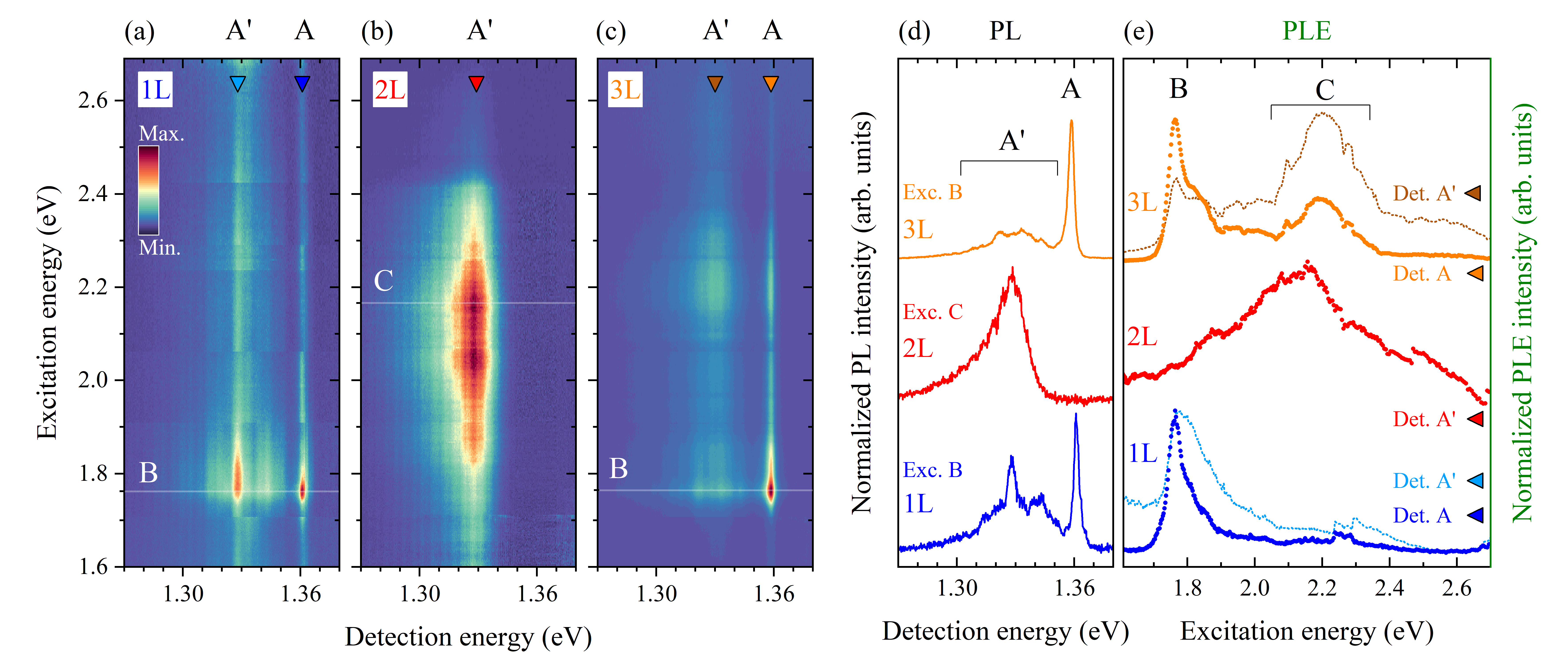}
    \caption
    {\label{fig2}
    False-color maps of the normalized PLE spectroscopy on (a) 1L, (b) 2L, (c) and 3L \crsbr.
    (d) PL spectra at selected excitation energies (whit lines) for the three systems: 1L and 3L excited at the B exciton ($E_{exc.}$=1.76~eV), and 2L excited at the C exciton ($E_{exc.}$=2.17~eV).
    (e) PLE spectra at detection energies A (1.36~eV) and A\textquotesingle~(1.33~eV) (indicated by triangles) for 1L, 2L (only at A\textquotesingle), and 3L \crsbr.
    Spectra in panels (d) and (e) are scaled and shifted vertically for clarity.
    }
\end{figure*}

For magnetic measurements, a millimeter size \crsbr~crystal was probed in a superconducting quantum interference device (SQUID).
Temperature-dependent magnetic susceptibility ($\upchi$) was probed in both zero-field-cooled (ZFC) and field-cooled-on-cooling (FCC) protocols, exhibithing a perfectly reversible behavior.
A detailed analysis is shown in the SI.
The intrinsic anisotropy of \crsbr~leads to different magnetic responses for measurements performed along each of the sample's crystallographic axes.
Namely, the $\hat{a}$, $\hat{b}$, and $\hat{c}$ axes correspond to the intermediate, easy, and hard magnetization axes, respectively~\cite{goser1990, fabrizio2022, francisco2023}.
Due to the needle-like shape of \crsbr, it is easy to identify the $\hat{a}$ axis, which coincides with the long edge of the needle.
Here, we focus on temperature-dependent $\upchi$ measured in the FCC mode under $B_\textrm{ex}$=0.01~T oriented parallel to the $\hat{a}$ and $\hat{b}$ sample axes (see Fig.~\ref{fig1}(c)). 
For temperatures above $T$=200~K, $\upchi$ approaches zero for both $B_\textrm{ex}$ orientations, confirming the presence of the paramagnetic phase.
Below this temperature, the magnetization reaches a maximum around 130~K, denoting the crossover to AFM ordering.
This is consistent with the previously reported T$_\textrm{N}$~\cite{telford2020, nathan2021}.
The easy axis configuration ($\Vec{B}\parallel\hat{b}$) shows a flattening in $\upchi$ for the AFM regime ($T$<130 K), showing that $B_\textrm{ex}$=0.01~T is not enough to magnetize \crsbr~in this direction.
However, along $\hat{a}$ direction, $B_\textrm{ex}$=0.01~T is enough to cant the spins and establish a net magnetic moment.
In fact, along the magnetic easy axis, the AFM ordering is stable up to $B_\textrm{ex}$=0.5~T, and higher $B_\textrm{ex}$ leads to spin flipping and increasing the $\upchi$~\cite{telford2020, sara2022}.
The analogous effect is not observed towards the intermediate axis, where $\upchi$ increases linearly with $B_\textrm{ex}$~\cite{telford2020, sara2022}.
The origin of the magnetic anisotropy is related to the different magnetic paths along $\hat{a}$ and $\hat{b}$, with the Br spin orbit coupling favoring the Cr spin alignment along $\hat{b}$~\cite{Huang2020, sara2022}.
Using the Curie-Weiss law, $\upchi=\upchi_{0}+C/(T-\uptheta_\textrm{CW})$, where $\upchi_{0}$ is a temperature-independent term, $C$ is the Curie constant and $\uptheta_\textrm{CW}$ is the Curie-Weiss temperature.
We determined $\uptheta_\textrm{CW}$=150~K for both configurations of magnetic field, which is lower than the previously reported values (164-185 K)~\cite{telford2020} and corresponds to the onset of local FM interaction in \crsbr, confirming the previous observations~\cite{telford2020, fabrizio2022, sara2022}.
Measurements at higher magnetic fields and isothermal magnetization are analyzed in the SI.


\begin{figure*}[]
    \centering
    \includegraphics[width=.87\linewidth]{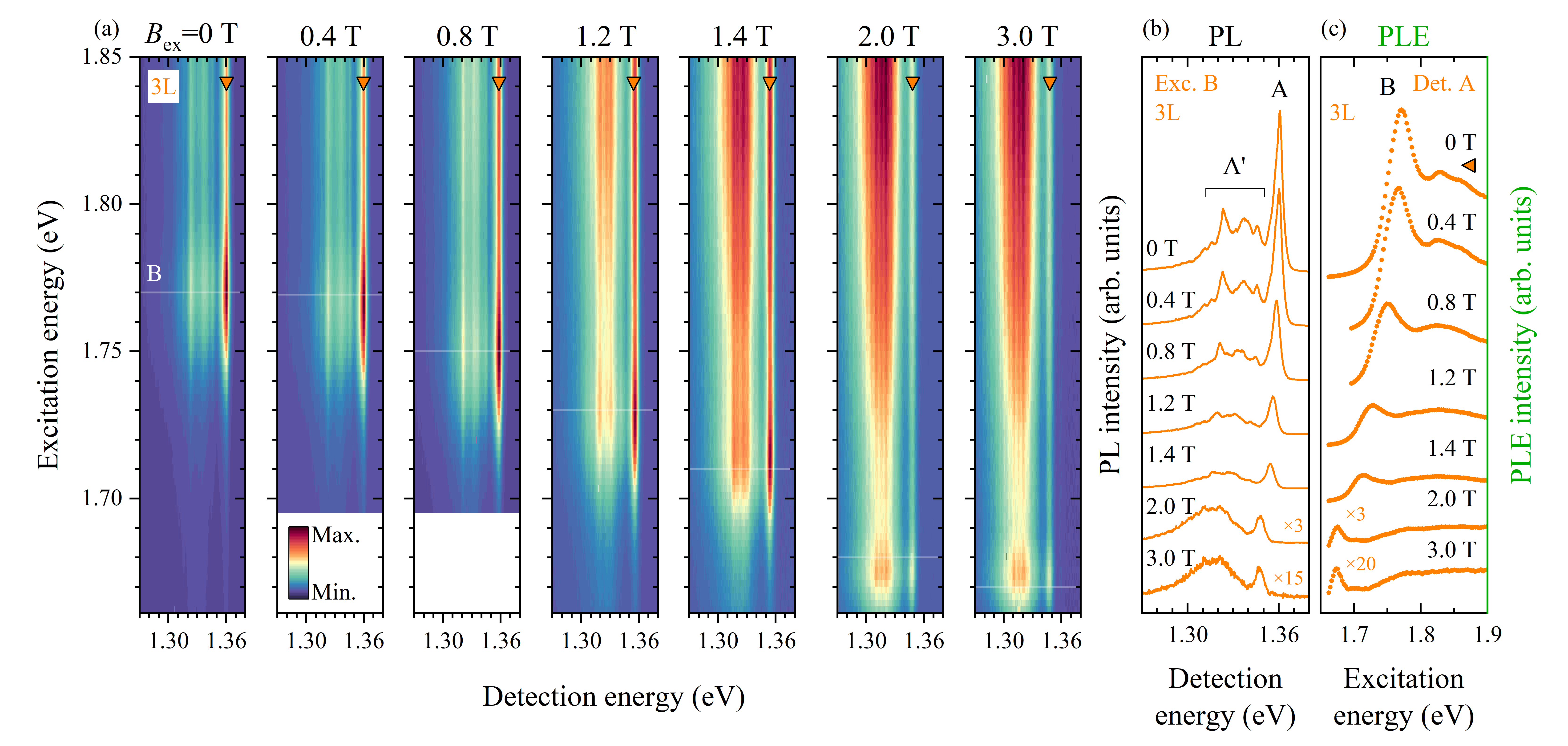}
    \caption
    {\label{fig3}
    (a) False-color map of the PLE spectroscopy measured on 3L \crsbr~around A exciton for selected values of the magnetic field (indicated on top of panels).
    For $B_\textrm{ex}$=0.4 T and 0.8 T, the excitation energy range starts from about 1.70~eV.
    Magnetic-evolutions of the (b) PL and (c) PLE spectra for 3L \crsbr~excited and detected correspondingly at the B (with lines) and A (triangles) excitons, respectively.
    The B and A energies evolve with the increase of the magnetic field.
    The spectra in panels (b) and (c) are scaled and shifted vertically for clarity.
    }
\end{figure*}

\subsection*{Emission and absorption spectra\\of few-layer \crsbr}

Figures~\ref{fig1}(d) and (e) show the low-temperature ($T$=10~K) PL and reflectance contrast (RC) spectra, respectively, measured on 1L, 2L, and 3L \crsbr.
Due to the strong in-plane anisotropy of the optical response~\cite{Klein2023}, the PL spectra were measured using an excitation lase of a 2.41 eV (515~nm) polarized along the $\hat{b}$ axis (see the Methods section for details).
The PL and RC spectra of the 1L and 3L samples are composed of two resonances: a high-energy narrow peak and a low-energy broad-band peak, labeled A and A\textquotesingle, respectively.
In contrast, the PL and RC spectra of 2L \crsbr~consists of a single broad band (A\textquotesingle).
The full width at half maximum of the A line is about 3~meV, whereas the one of A\textquotesingle~reaches values of the order of 25~meV.
Our results of the optical response of 1L, 2L, and 3L \crsbr~agree very well with literature~\cite{nathan2021, francisco2023, Klein2023}, confirming the proposed flake thickness assignment. 

The most important question is the origin of the observed A and A\textquotesingle~transitions, which is still being debated in the community~\cite{nathan2021, Klein2023, francisco2023, Komar2024, Shao2025, Smiertka2025}.
The first works~\cite{nathan2021, Klein2023} reported that the A transition is related to Wannier-Mott excitons formed in the vicinity of the CrSBr band gap, similarly to free excitons in transition metal dichalcogenides~\cite{Koperski2017, Molas2019}.
However, \crsbr~excitons are highly delocalized in reciprocal space, leading to a well confined excitons along the $\hat{c}$- and $\hat{b}$-axis, giving rise to a quasi-1D character~\cite{Wu2022, Shao2025}.
Recent calculations further indicate a significant binding energy of several hundred meV~\cite{Watson2024, Smiertka2025}.
In fact this value is smaller than the one observed in pure Frenkel excitons, as in CrX$_3$~\cite{Grzeszczyk2023}, but confers to A/A\textquotesingle~\crsbr~excitons a Frenkel character which cannot be omitted.
Y. Shao $et~al.$~\cite{Shao2025} proposed that the A and A\textquotesingle~transitions can be ascribed to the surface and bulk excitons, respectively.
This scenario proposes that the A resonance should not be observed in the 1L and 2L limits.
While for the 2L, only the A\textquotesingle~is reported in the literature~\cite{nathan2021, Klein2023, francisco2023, Shao2025}, the 1L PL spectra are composed of either a single A\textquotesingle~peak ~\cite{nathan2021, Shao2025} or both A and A\textquotesingle~peaks~\cite{Klein2023, francisco2023}.
In our case, the latter pattern is apparent.
This discrepancy is not clear, $e.g.$ it has been proposed that the A intensity can be enhanced by the ionization of impurities~\cite{francisco2023}, but we will demonstrate in the following that the relative intensity of the A and A\textquotesingle~lines can be highly tuned by the excitation energy.
RC spectra were also recorded in a wider energy range of 1.6~eV to 2.8 eV (Figure~\ref{fig1}(e) and SI for more details).
The low-energy A and A\textquotesingle~peaks are well resolved, matching PL data.
However, higher-energy transitions exhibited low intensity, limiting a more precise analysis.
Therefore, we focus on excitation-dependent PL measurements, $i.e.$ PLE.


PLE experiments were carried out on 1L, 2L, and 3L \crsbr~(Fig.~\ref{fig2}).
The PLE technique provides a quasi-absorption spectrum, similar to RC measurements.
However, PLE resonances correspond to excited states of the monitored PL emission, while RC transitions are associated with all absorption features in the studied material~\cite{Grzeszczyk2021, Karmakar2023}.
Figs.~\ref{fig2}(a)-(c) show the normalized low-temperature ($T$=10~K) PLE maps detected on 1L, 2L, and 3L samples, respectively.
Horizontal cuts along the B and C excitonic transitions, $i.e.$ at the excitation energies of 1.76~eV and 2.17~eV, yeld the PL spectra presented in Fig.~\ref{fig2}(d). 
Although the 2L emission spectrum (Fig.~\ref{fig2}(d)) closely resembles that in Fig.~\ref{fig1}(d), this is not the case for the 1L and 3L spectra, despite both being basically composed of the same A and A\textquotesingle emissions.
Firstly, under B excitation, the A intensity is improved, exceeding A\textquotesingle, opposite to what was observed in Fig.~\ref{fig1}(d).
Secondly, the A\textquotesingle~shapes modify from the broad bands (Fig.~\ref{fig1}(d)) to those formed by sets of a few individual emission lines in Fig.~\ref{fig2}(d).
The excitation-energy PL evolution is shown in analyzed in SI for 1L to 3L samples.
Indeed, as the excitation energy decreases toward the B resonance, the 1L/3L emission spectra evolve from those in Fig.~\ref{fig1}(d) to those in Fig.~\ref{fig2}(d), and finally resemble the initial spectra again at lower excitation energies, around 1.6~eV.
Thus, our PLE results for 1L indicate that the A and A\textquotesingle~intensity can be strongly adjusted by tuning the excitation energy.
This finding likely resolves the discrepancy in the presence of the A emission in the PL spectra of 1L \crsbr~reported by the community~\cite{nathan2021, Klein2023, francisco2023, Shao2025}.
The variation of A\textquotesingle~shape in the 1L/3L layers suggests that these emissions can be strongly quenched by resonant excitation through the B exciton, suppressing non-resonant scattering, we conjecture that such phenomenon occurs via a series of phonon emissions, apparent under excitation with larger and smaller energies.
Furthermore, the origin of the A\textquotesingle~bands in the 2L is different from the one observed in 1L/3L, as denoted by its complete absence of influence or the strong effect of the excitation energy.

\begin{figure}[ht]
    \centering
    \includegraphics[width=1\linewidth]{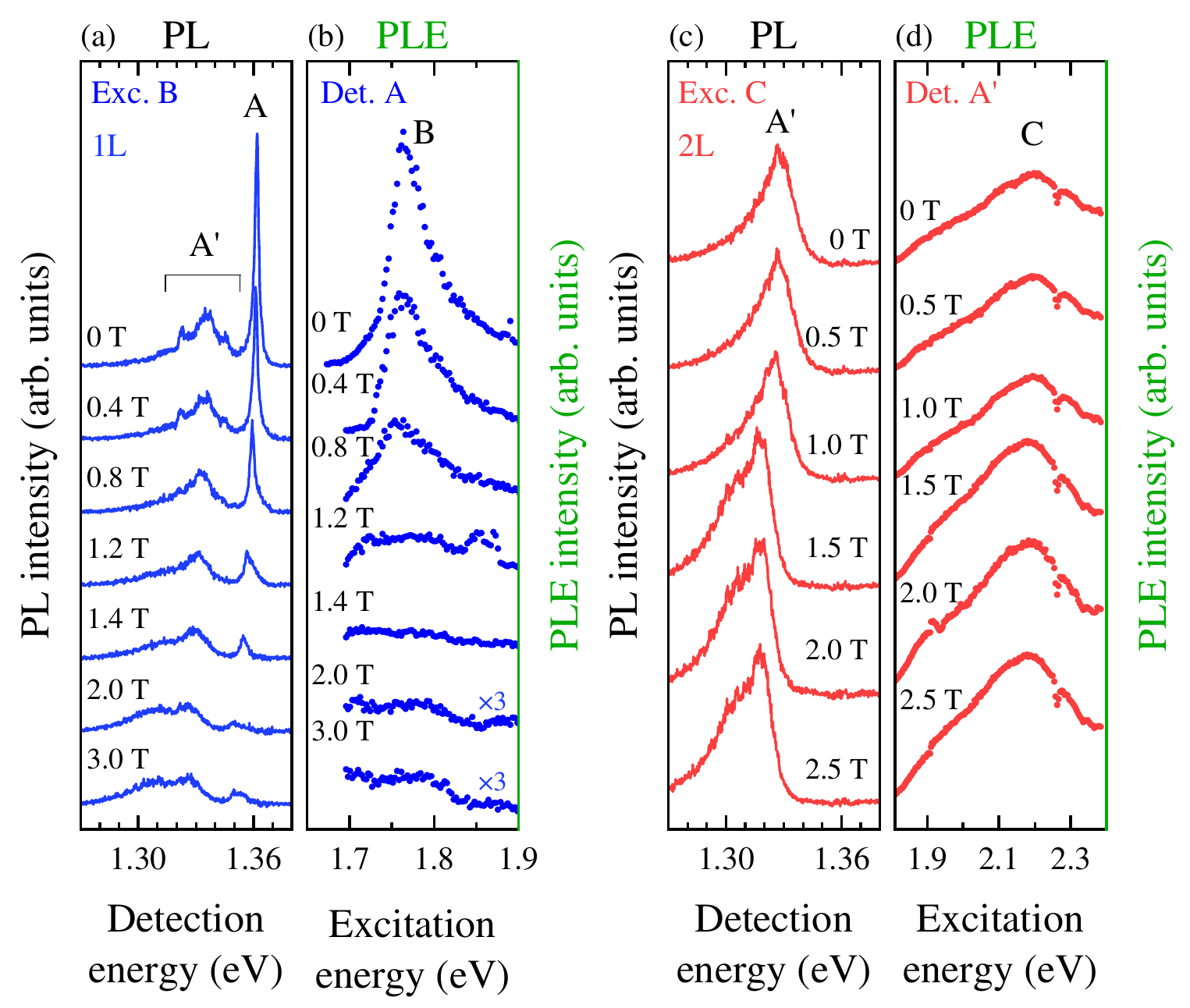}
    \caption
    {\label{fig4}
    Magnetic-evolution of PL (a), (c) and PLE (b), (d) for 1L and 2L \crsbr~excited and detected correspondingly at the B amd C excitons.
    The spectra are scaled and shifted vertically for clarity.
    }
\end{figure}

To investigate the excited states of the A and A\textquotesingle~emissions, we integrated intensities of 20 points around resonances, marked by the colored triangles in Figs.~\ref{fig2}(a)-(c), and plot corresponding PLE spectra detected on the A and A\textquotesingle~transitions in Fig.~\ref{fig2}(e).
The excitation spectra obtained for 1L and 3L are substantially different.
The 3L PLE spectrum consists of two resonances, labeled B and C, apparent correspondingly at about 1.75~eV and 2.2~eV.
The B (C) linewidth are approximately 100 (200)~meV, significantly broader than the A linewidth, which is on the order of 3~meV.
The band origin of the B exciton, however, is very similar to that of A, albeit involving a different set of bands~\cite{Smiertka2025}.
Moreover, the calculated spatial extent of the B exciton along the $\hat{b}$-axis compared to the A exciton is almost four times larger~\cite{Smiertka2025}, suggesting much smaller binding energy of the B excitonic state versus the A binding energy. 
Additionally, B transition in 3L is composed of two resonances that come from different subband spectral compositions in conduction and valence band, similar to the bulk case~\cite{Komar2024, Smiertka2025}.

According to a recent calculation~\cite{Bianchi2023, Smiertka2025}, \crsbr~presents an energy gap of 2.07 eV, which implies a Frenkel-like behavior for A exciton with a relatively small binding energy (0.7 eV).
The B exciton binding energy is approximately 0.3 eV and displays more pronounced Wannier-Mott characteristics~\cite{Klein2023, Shao2025, Smiertka2025}.
The coexistence of both Frenkel and Wannier excitons in the same material is highly uncommon in layered systems.
The 1L \crsbr~PLE spectrum is formed by a single peak of the B transition (Fig.~\ref{fig2}(e)), similar to the 3L flake, but the absence of C highlights the difference between the two thicknesses.
Note that this discrepancy between 1L and 3L may facilitate the identification of the number of layers in CrSBr.

Similarly to the significant distinction between the PL of the 2L and 1L/3L flakes, the excitation spectrum of 2L is formed by a single, 340 meV broad C feature (Fig.~\ref{fig2}(e)).
We use the same notation for the C resonances in 2L and 3L due to their similar energies, but the origin probably can be different, analogously to the aforementioned A\textquotesingle~emissions in 2L and 1L/3L limits.
However, the assigning of the C exciton is beyond the scope of our experimental work.


\subsection*{Magneto-Excitonic Dynamics\\in Few-Layer \crsbr}

We now turn our attention to the AFM to FM transition in few-layer \crsbr.
We employ magneto-PLE technique to investigate the effect of the out-of-plane $B_\textrm{ex}$ on the optical properties of few-layer \crsbr~at $T$=10~K.
The field orientation corresponds to the material's hard magnetization axis ($\hat{c}$).
Fig.~\ref{fig3}(a) shows a set of PLE maps measured on 3L \crsbr~under different $B_\textrm{ex}$ (indicated on top of each panel) excited in the vicinity of the B transition.
The 3L magneto-evolution PL (Fig.~\ref{fig3}(b)) and PLE (Fig.~\ref{fig3}(c)) spectra were captured at B and A transitions, respectively.
The increase in the $B_\textrm{ex}$ leads to a redshift of both the A and A\textquotesingle~peak positions and a significant change in their relative intensities.
At the same time, the A\textquotesingle~shape changes from well-resolved emission lines at 0~T to a broad-band emission at 3~T.
Here, it is worthy to notice that 0~T spectra in Figs.~\ref{fig2} and~\ref{fig3} may vary slightly, as they were measured using distinct experimental setups.
Interestingly, the FM ordering of the 3L (Fig.~\ref{fig3}(b), 3.0~T spectrum) causes on PL spectra a similar effect as the non-resonant excitation (Fig.~\ref{fig2}(c) and SI), suggesting a similar effect on the excitonic states.
The PLE spectra of the A exciton mirrored the behavior of its corresponding PL, $i.e.$ B peak redshifts and vanishes at high applied magnetic fields.
However, the B peak redshift is much larger than that of the A peak, and its origin is discussed in the following paragraphs.

\begin{figure}[]
    \centering
    \includegraphics[width=.9\linewidth]{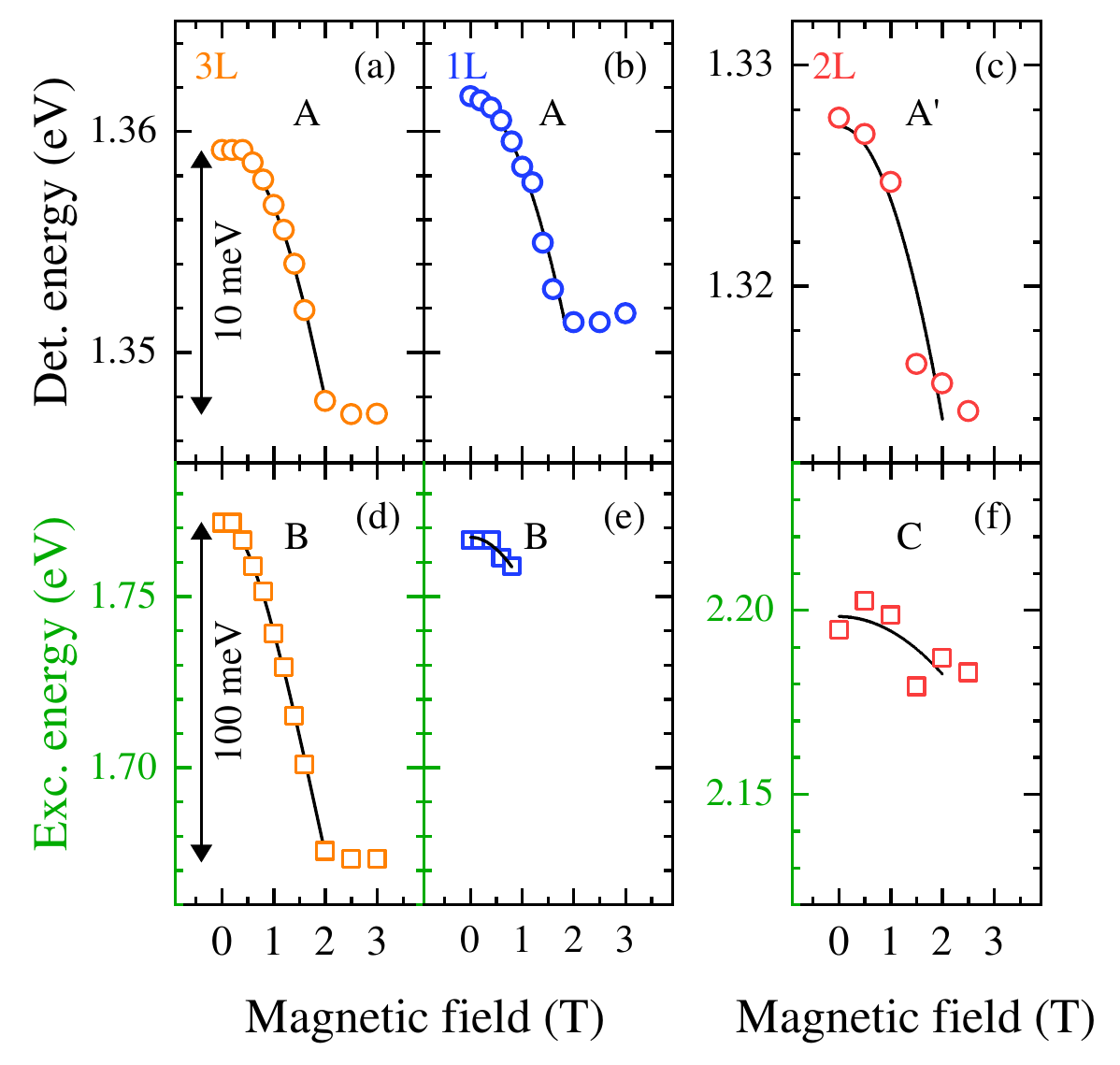}
    \caption
    {\label{fig5}
    Magnetic field dependences of the A, for 3L/1L (a)/(b), and A\textquotesingle, for 2L (c), obtained from the PL spectra.
    Magnetic field dependences of the B, for 3L/1L (d)/(e), and C, for 2L (f), obtained from the PLE spectra. 
    The solid curves represent fits according to the equation described in the text.
    }
\end{figure}

For a comprehensive analysis, we also study the magnetic field evolution of the PL and PLE spectra of 1L and 2L \crsbr.
The PLE maps of the 1L and 2L samples are provided in the SI. 
For the 1L flake, the $B_\textrm{ex}$ causes the A, A\textquotesingle, and B peaks to redshift and significantly quenches their relative intensities (Fig.~\ref{fig4}(a)-(b)), similar to the 3L case (Fig.~\ref{fig3}(b)-(c)).
As well as in 3L case, the A\textquotesingle~shape varies from the set of well-resolved emission lines to a broad band, however, the B resonance disappears completely from the PLE spectra at about 1~T.
The analogous changes induced by the magnetic field in 1L and 3L suggest that the properties of the A, A\textquotesingle, and B excitons share similar properties.
This is consistent with the fact that 3L can be magnetically regarded as a 2L (fully AFM) plus an additional 1L.
The strong disappearance of the A, A\textquotesingle, and B intensities can be described in terms of the lower oscillator strength of a given transition~\cite{Schuller2002, krelle2025}.
This suggests a smaller excitonic binding energy and a correspondingly larger exciton radius.
These effects are associated to the change in magnetic ordering transition from AFM to FM induced by $B_\textrm{ex}$~\cite{nathan2021, Klein2022, Komar2024, Smiertka2025}.
The PL and PLE results obtained for 2L \crsbr, shown in Figs.~\ref{fig4}(c) and (d), are completely different compared to the aforementioned 1L and 3L cases.
The intensities and shape of the A\textquotesingle~and C resonances are barely affected by $B_\textrm{ex}$. However, the peak positions show a redshift with increasing $B_\textrm{ex}$.
This suggests that the excitonic nature of the A\textquotesingle~and C resonances in 2L has distinct characteristics compared to those observed in 1L and 3L flakes.
At this point we can speculate that these differences originate from the behavior of the in-plane magnet moment.
The 2L \crsbr~has always zero in-plane net magnetic moment, independent of external field.
In contrast, 1L and 3L exhibit a finite in-plane net magnet moment due to a single \crsbr~layer.
The application of an out-of-plane $B_\textrm{ex}$ in 1L/3L can therefore be correlated with a reduction of the in-plane net magnetic moment in these systems.

Finally, we determine the magnitude of the magnetic field-induced shifts of A, A\textquotesingle~, B, and C excitonic transitions of 1L to 3L \crsbr.
First, we focus on the transition observed in the vicinity of the \crsbr~optical band gap.
The magnetic field evolution of A for 3L and 1L is summarized in Figs.~\ref{fig5}(a) and (b), respectively, and that of A\textquotesingle~for 2L is shown in Fig.~\ref{fig5}(c).
The extracted shifts of the A and A\textquotesingle~energies are around 10~meV for all investigated layers at $B_c$.
These results agree very well with the values previously reported for thin and bulk CrSBr flakes~\cite{nathan2021, Klein2022, tabataba2024, Komar2024, Christin2025, Smiertka2025}.
Excitonic transitions at higher energies are described by different magnetic-field-induced evolutions in the thin \crsbr~layers studied: (i) for 3L, we observe the redshift of the B exciton, reaching a value of around 100~meV. 
Such a giant magnetic dependence has been reported for \crsbr~bulk~\cite{Komar2024, Smiertka2025}. 
Komar $et~al.$~\cite{Komar2024} described it as the hybridization of excitons between layers, which is suppressed in the AFM phase but enhanced in the FM phase.
Watson $et~al.$~\cite{Watson2024} calculated the band gap of \crsbr~to be 2.07 eV, which decreases by 110 meV under FM ordering.
\'Smiertka $et~al.$~\cite{Smiertka2025} observed that the B exciton follows the conduction band reduction, exhibiting a Wannier-Mott character.
Meanwhile, the A and A\textquotesingle~exciton, display predominantly a Frenkel-like behavior, with a weaker dependence on the host band structure.
(ii) In the case of 1L, the B exciton could only be traced up to 1~T, with the magnetic-field-induced redshift of approx. 10~meV.
This energy modification is slightly smaller than that observed in the same range for 3L \crsbr, but it exceeds the redshift of the A and A\textquotesingle~excitons over the entire $B_\textrm{ex}$ range (0~T to 3~T).
(iii) The 2L C exciton shifts by roughly 20~meV up to the $B_c$.
However, 2L \crsbr~exhibits an unique excitonic response~\cite{Dirnberger2023, Dirnberger2025}, and further studies are needed to understand its underlying mechanisms.

The excitonic shifts were extracted with the usual quadratic dependence of energy, which can be described by $E(B_\textrm{ex})=E_0-tB_\textrm{ex}^2$, where $E_0$ is the exciton energy and $t$ is related to both the interlayer hole tunneling and the energy splitting between the intralayer and interlayer excitons~\cite{tabataba2024, Christin2025}.
Most of data are well described by the quadratic dependence of energy model and the fitting parameters can be found in SI.
Even with quadratic aspect, the 3L B resonance (fig.~\ref{fig5}~(d)) deviates substantially from the quadratic dependence.
To describe such huge magnetic dependence, we apply the phenomenological approach suggested by Komar $et~al.$~\cite{Komar2024}.
More details can be found in SI.

\subsection*{Conclusions \label{sec:Conclusions}}

In summary, we have studied 1L, 2L, and 3L \crsbr~PL and PLE spectra.
We observed that the A exciton in the 1L and 3L systems can be tuned by the excitation energy, under-reported by earlier studies.
The PLE spectra revealed two prominent resonant energies: first one around 1.77~eV for 1L and 3L, associated with the B exciton. The other at 2.17~eV for 2L. 
Furthermore, the PLE spectrum for 3L exhibits an additional resonance around 2.20~eV, which is not present in 1L.
This extra feature provides a spectral distinction between 1L and 3L CrSBr.
Under $B_\textrm{ex}$ (applied in $c$-axis) the excitonic features redshift up to $B_c$.
However, while the A/A\textquotesingle~shift is about 10~meV, the 3L B exciton exhibits a giant dependence of around 100~meV.
This dual behavior in 3L CrSBr is attributed to an atypical property of \crsbr, which supports both Frenkel- (A/A\textquotesingle) and Wannier-Mott-like (B) excitons.
These effects were confirmed to persist even in few-layer systems, highlighting the robustness of \crsbr~magneto-excitonic coupling.
The 2L system shows a singular broad excitonic response and a moderate magnetic shift of the C exciton.
Our study on \crsbr~challenges the conventional separation of Frenkel and Wannier excitons. Thus, opening new opportunities to study hybrid excitonic behavior in 2D materials.


\section*{Methods\label{sec:Methods}}
\textbf{CrSBr crystal synthesis.}
For synthesis, Chromium (99.99$\%$, -60 mesh, Chemsavers, USA), bromine (99.9999$\%$, Sigma-Aldrich, Czech Republic) and sulfur (99.9999$\%$, Stanford Materials, USA) were mixed in a stoichiometric ratio in a quartz ampule 40x200 mm) corresponding to 15 g of \crsbr. 
Bromine excess of 0.5 g was used to enhance vapor transport.
The material was pre-reacted in an ampule using a crucible furnace at 500$^\circ$C for 50 hours, while the second end of the ampule was kept below 150$^\circ$C.
The heating procedure was repeated two times until the liquid bromine disappeared.
The ampule was placed in a horizontal two-zone furnace for crystal growth.
First, the growth zone was heated to 900$^\circ$C, while the source zone was heated to 700$^\circ$C for 2 days.
For the growth, the thermal gradient was reversed and the source zone was heated to 900$^\circ$C and the growth zone to 800$^\circ$C over a period of 10 days.
The crystals with dimensions up to 2x15 mm were removed from the ampule in an Ar glovebox.

\textbf{Samples fabrication.}
1L to 3L \crsbr~flakes were obtained on a degenerately doped (320 nm)SiO$_2$/Si substrate by polydimethylsiloxane-based exfoliation~\cite{gomez2014}.
The flakes of interest were first identified by visual inspection under an optical microscope and then subjected to atomic force microscopy characterization to unambiguously determine their thicknesses.

\textbf{Magnetometry measurements.} Magnetometry measurements were performed using a SQUID magnetometer on a \crsbr~ bulk crystal, with approximate dimensions 3 mm × 1 mm × 1 mm and a mass of 10.59 mg. 
The magnetic moment of the sample was measured in the temperature range 2~K < T < 300~K, and magnetic fields up to 7~T.
Two measurements protocol were used: Zero-field-Cooled (ZFC: The sample is cooled in the absence of magnetic fields, and measured upon warming) and Field-Cooled-on-Cooling (FCC: The sample is measured upon cooling).
Measurements are performed in two crystallographic configurations: $\Vec{B} \parallel \hat{a}$ and $\Vec{B} \parallel \hat{b}$.

\textbf{Magneto-optical spectroscopy.} The PL and RC experiments were performed using a $\lambda$=515 nm (2.41~eV) continuous wave (CW) laser diode and a 100 W tungsten halogen lamp, respectively.
PLE experiments were performed at $T$=10 K using a supercontinuum light source coupled with a monochromator as an excitation source, which works as a tunable excitation light.
As the light from the tunable source was not constant along the whole wavelength range, thus, to avoid artificial effects, the PLE investigations were acquired in small excitation ranges, and discontinuities can be seen in Fig.~\ref{fig2}.

The studied samples were placed on a cold finger in a continuous-flow cryostat mounted on x-y motorized positioners. 
The excitation light was focused by means of a 50x long-working distance objective with a 0.55 numerical aperture (NA) that produced a spot of about 1 $\upmu$m diameter in PL/PLE measurements and 4 $\upmu$m diameter in RC measurements. 
The signal was collected via the same microscope objective, sent through a 0.75 m monochromator, and then detected using a charge-coupled device (CCD) camera. 

Magneto-optical experiments were performed in the Faraday configuration using a free-beam-optics arrangement in a superconducting coil producing magnetic fields up to 16 T. 
The investigated samples were placed on top of an x-y-z piezoelectric stage in a helium gas atmosphere at the temperature of 10~K. 
The excitation light was focused with a 100x microscope objective with NA=0.82 on a spot of about 1 $\upmu$m diameter.
The signal was collected through the same microscope objective, sent through a 0.50~m monochromator, and then detected using a CCD camera. The excitation power focused on the sample was kept below 100~$\upmu$W during all measurements to avoid local heating.
\section*{Supplementary Information}
Supplementary Information contains additional results of the atomic force microscopy, magnetic susceptibility, RC spectra, excitation-dependent PL spectra, and magnetic field evolution of the PLE maps.  

\section*{Data availability statement}
All the data necessary to conclude this study, are presented in the manuscript and SI. The raw data are available from the corresponding authors upon reasonable request.

\section*{Acknowledgments}
The work was supported by the National Science Centre, Poland (Grant No. 2020/37/B/ST3/02311) and as part of a project co-financed by the Polish Ministry of Science and Higher Education under contract no. 2025/WK/01.
Z.S. was supported by ERC-CZ program (project LL2101) from Ministry of Education Youth and Sports (MEYS) and  by the project Advanced Functional Nanorobots (reg. No. CZ.02.1.01/0.0/0.0/15$\_$003/0000444 financed by the EFRR).


\bibliographystyle{apsrev4-2}
\bibliography{biblio}
\end{document}


\title{SUPPLEMENTARY INFORMATION:\\Magneto-Excitonic Duality From Monolayer to Trilayer \crsbr}


\def \FUW{University of Warsaw, Faculty of Physics, Pasteura 5, 02-093 Warsaw, Poland}
\def \UCT{Department of Inorganic Chemistry, University of Chemistry and Technology, Technická 5, 166 28 Prague 6, Prague, Czech Republic}

\author{Igor Antoniazzi} 
\email{igor.antoniazzi@fuw.edu.pl}
\affiliation{\FUW}
\author{\L{}ucja Kipczak}
\affiliation{\FUW}
\author{Bruno Camargo}
\affiliation{\FUW}
\author{Gayatri}
\affiliation{\FUW}
\author{Chinmay~Mohanty}
\affiliation{\FUW}
\author{Kseniia~Mosina}
\affiliation{\UCT}
\author{Zden\v{e}k Sofer}
\affiliation{\UCT}
\author{Adam Babi\'nski}
\affiliation{\FUW}
\author{Arka Karmakar}
\email{arka.karmakar@fuw.edu.pl}
\affiliation{\FUW}
\author{Maciej R. Molas}
\email{maciej.molas@fuw.edu.pl}
\affiliation{\FUW}

\begin{abstract} 

\end{abstract}

\maketitle

\subsection{Atomic force microscopy}

\begin{figure}[h]
    \centering
    \includegraphics[width=.5\linewidth]{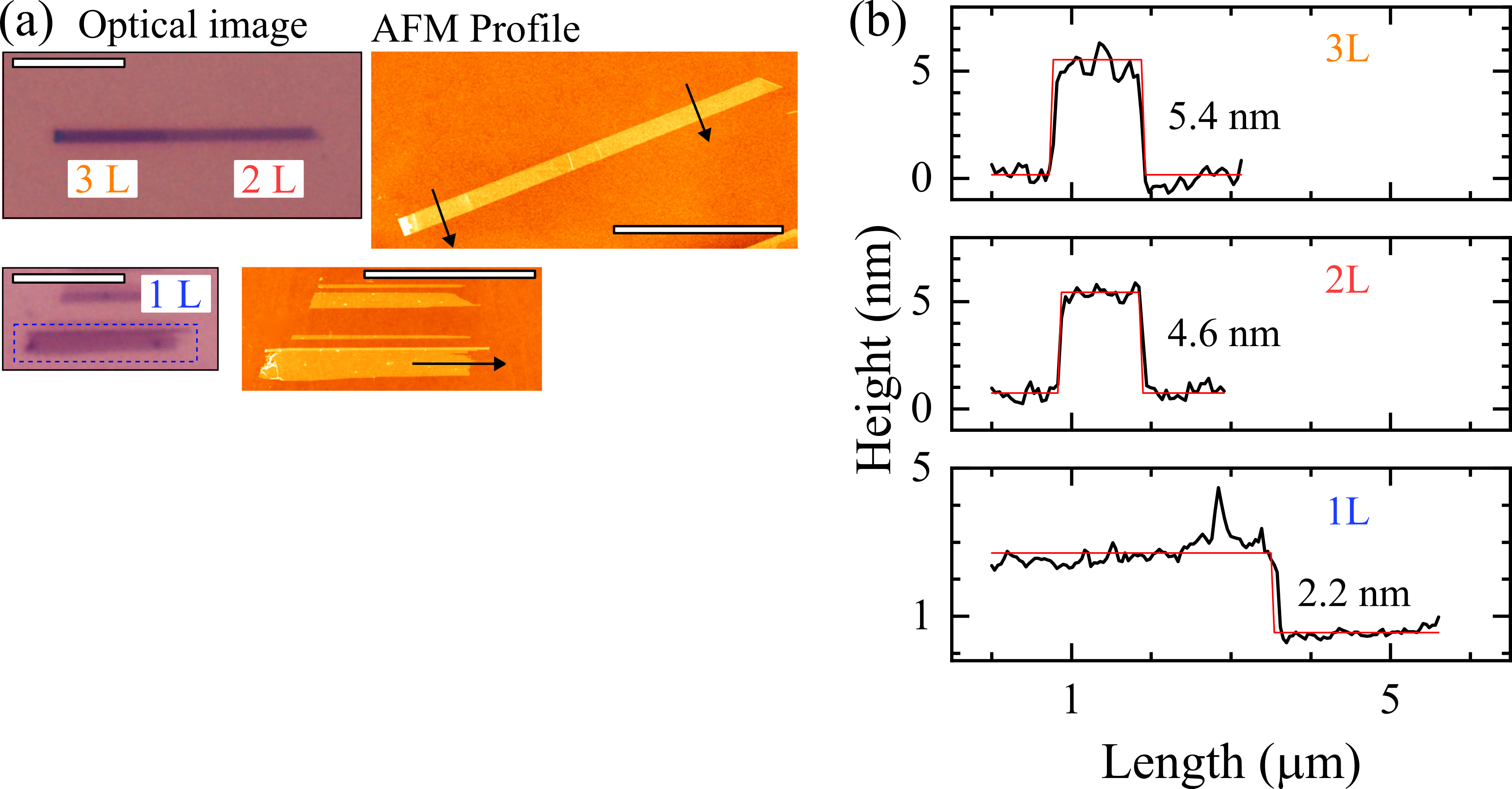}
    \caption{\label{si: figs. 1}
    (a) Optical image and afm of 1L, 2L, and 3L \crsbr.
    The white bar has 10~$\upmu$m.
    (b) The afm profiles extracted over the black arrows in (a) as well as the step function fitting (red line).
    }
\end{figure}

The few-layers \crsbr~were obtained by the well established exfoliation method.
Flakes of \crsbr~were exfoliated on polydimethylsiloxane (PDMS) and after transferred to Si/SiO$_2$.
The optical image as well as the atomic force microscopy (afm) image of the layers are displayed in Fig.~\ref{si: figs. 1}~a.
The profiles, measured over the black arrows are shown in Fig.~\ref{si: figs. 1}~b together a step function fitting (red line).
\crsbr~has $c$=0.7965 nm~\cite{goser1990}, thereby, we believe that background, as well as the exfoliation processes, are increasing the thickness of our samples, as it was observed to other 2D materials.~\cite{Blake2007, zhihong2007, Casiraghi2009, bertolazzi2013, junjie2015}
We estimated an interfacial layer of 1.4 nm and 3 nm for 1L, and 2L (3L), respectively.

\clearpage

\subsection{Magnetic susceptibility}

\begin{figure}[h]
    \centering
    \includegraphics[width=.9\linewidth]{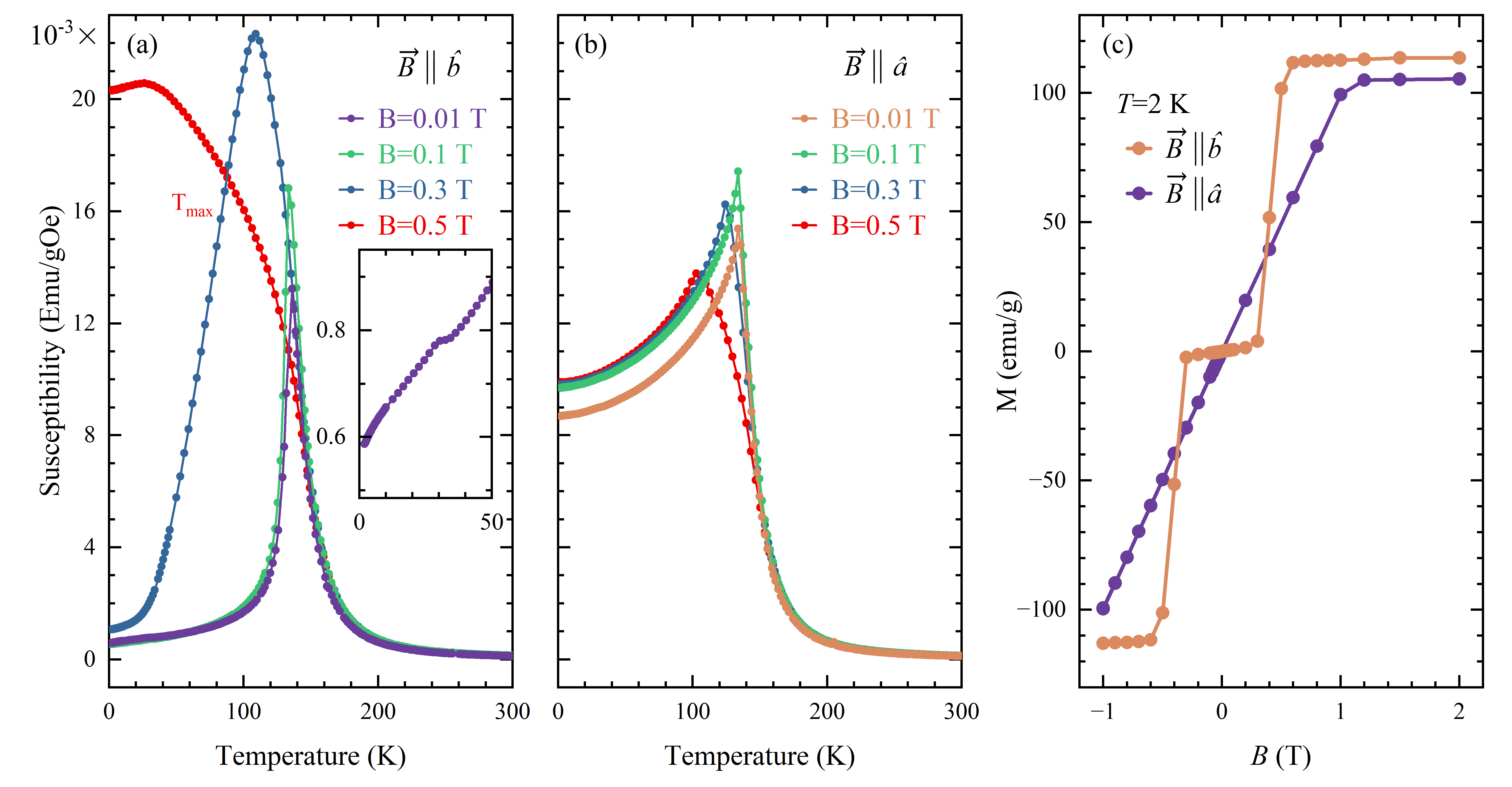}
    \caption{\label{si: figs. 2}
    The $\upchi$ measured at $\Vec{B} \bot \hat{a}$ (a) and $\Vec{B} \parallel \hat{a}$ (b) for different $B$.
    (c) The M curves for a \crsbr~single crystal at $\Vec{B} \parallel \hat{a}$ (purple) and $\Vec{B} \bot \hat{a}$ (orange) are shown for $T$=2~K.
    }
\end{figure}

Magnetic susceptibility ($\upchi$) measured perpendicular ($\Vec{B} \bot \hat{a}$) (Fig.~\ref{si: figs. 2}~a) and parallel ($\Vec{B} \parallel \hat{a}$) (Fig.~\ref{si: figs. 2}~b) to the sample's $\hat{a}$-axis for different $B$.
The inset shows the usual kink observed at $T\approx$30~K, quoted as a hidden-order transition, measured at 0.01~T.
Its origin is ascribed by Paz $et~al.$~\cite{sara2022} to a continuous reorientation of the internal magnetic field, leading to a spin-freezing and consequently to a spin dimensionality crossover (d=2 to d=1).
However, other authors described the hidden order transition in terms of an ordering of magnetic defects or incoherently coupled 1D electronic chains.~\cite{telford2020, Wu2022, constant2022}

The isothermal magnetization (M) loop performed at $T$=2~K revealed no signature of coercive fields (Fig.~\ref{si: figs. 2}~c).
For $\Vec{B} \parallel \hat{a}$ (purple) the magnetization increases linearly with $B$, saturating close to 1.1~T, while $\Vec{B} \bot \hat{a}$ (orange) revealed the presence of a sharp transition around fields of 0.05~T, suggesting an in-plane spin-flip transition.
The different saturation $B$ along the different crystal orientations, shows the magnetic anisotropyof \crsbr.~\cite{sara2022} 

\clearpage

\subsection{Reflectance contrast spectra}

The RC at high energy range indicates a the transition B at roughly 1.8~eV for 1L and 3L.
Close to 2.1~eV there is also C transition clearly observed for 2L and 3L, that is absent (or very weak) in 1L case.

\begin{figure}[h]
    \centering
    \includegraphics[width=.31\linewidth]{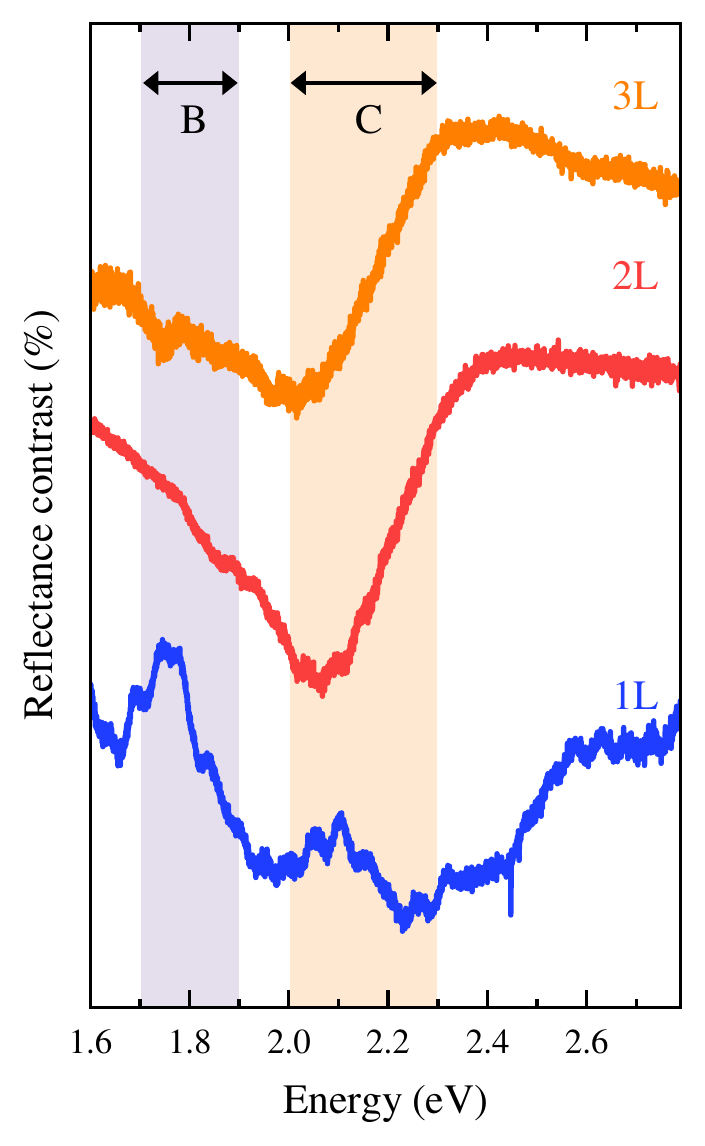}
    \caption{\label{si: figs. 3}
    The RC spectra for 1L, 2L, and 3L \crsbr~at high energy range.
    }
\end{figure}

\clearpage

\subsection{Energy-dependent photoluminescence spectra}

\begin{figure}[h]
    \centering
    \includegraphics[width=.65\linewidth]{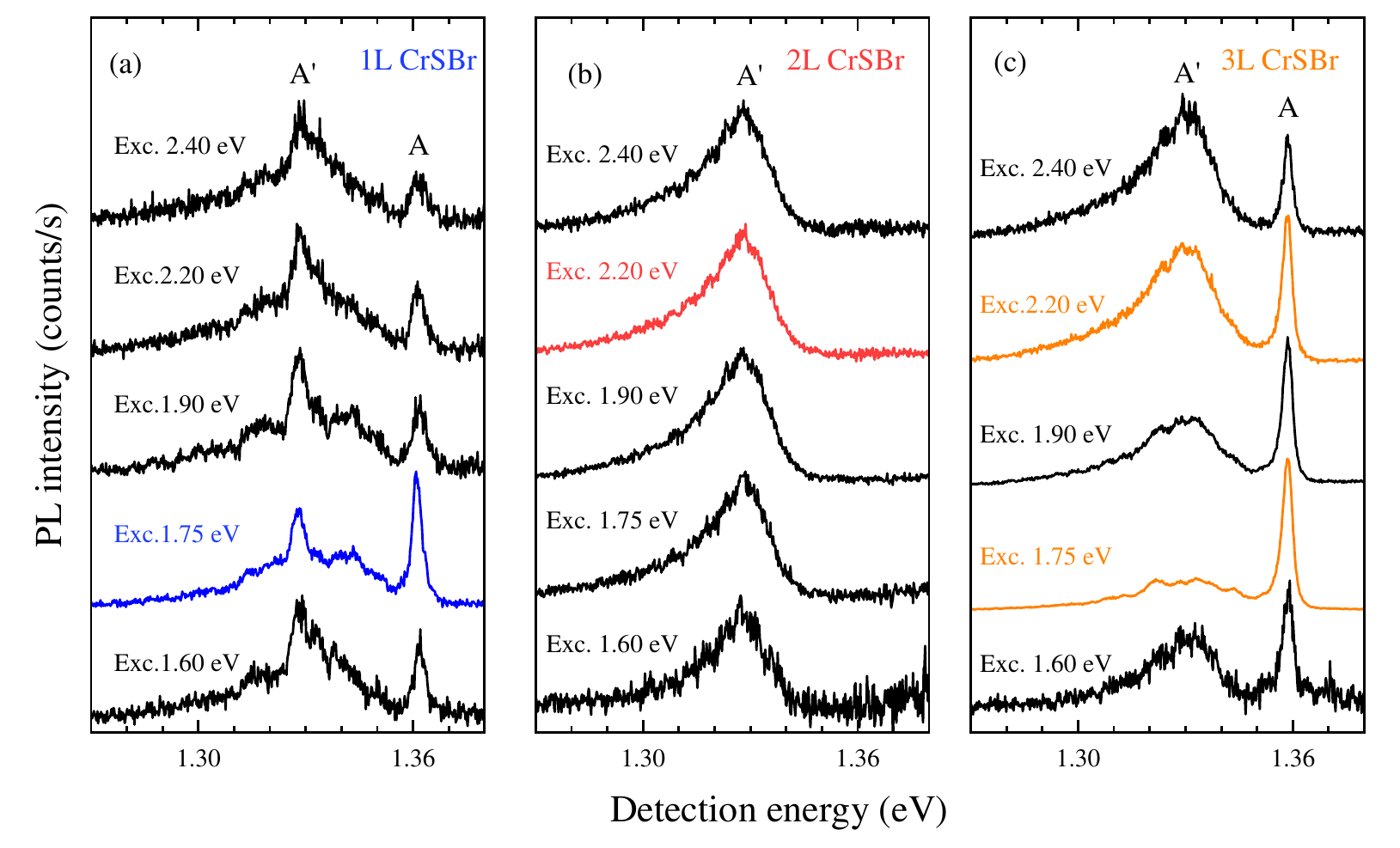}
    \caption{\label{si: figs. 4}
    Selected PLE spectra showing the evolution of 1L (a), 2L (b), and 3L (c) \crsbr~in respect of excitation increasing.
    }
\end{figure}

Selected PL spectra from the PLE maps (Fig. 2(a)-(c)) are shown in Fig.~\ref{si: figs. 4}(a)-(c) for 1L to 3L \crsbr, respectively.
Horizontal cuts at excitation energies 1.60 eV, 1.75 eV, 1.90 eV, 2.20 eV, and 2.40 eV reveal the A and A\textquotesingle~dynamics.
For 1L and 3L, an inversion in intensity height of A and A\textquotesingle~is observed, along with a set of individual emissions in the A\textquotesingle~range.
The A and A\textquotesingle~dependence gives an insight into why A is sometimes absent in the 1L system and shows that 2L A\textquotesingle has a different origin from the one observed in 1L/3L, not being affected by the B resonance excitation.

\clearpage

\subsection{Magnetic evolution of the photoluminescence excitation maps for 1L and 2L \crsbr}
\begin{figure}[h]
    \centering
    \includegraphics[width=.65\linewidth]{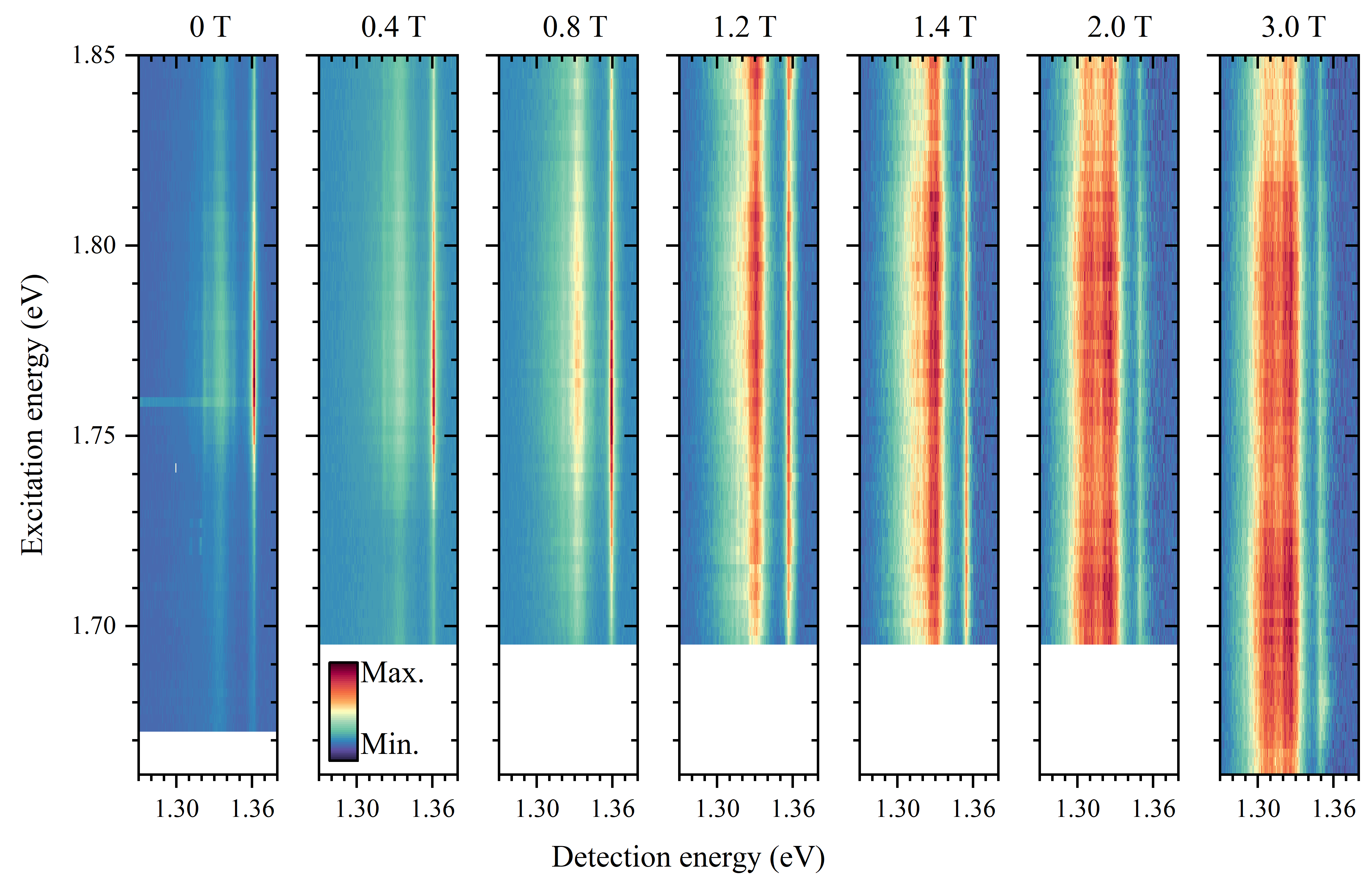}
    \caption{\label{si: figs. 5}
    Magnetic evolution PLE for 1L \crsbr.
    }
\end{figure}

\begin{figure}[h]
    \centering
    \includegraphics[width=.65\linewidth]{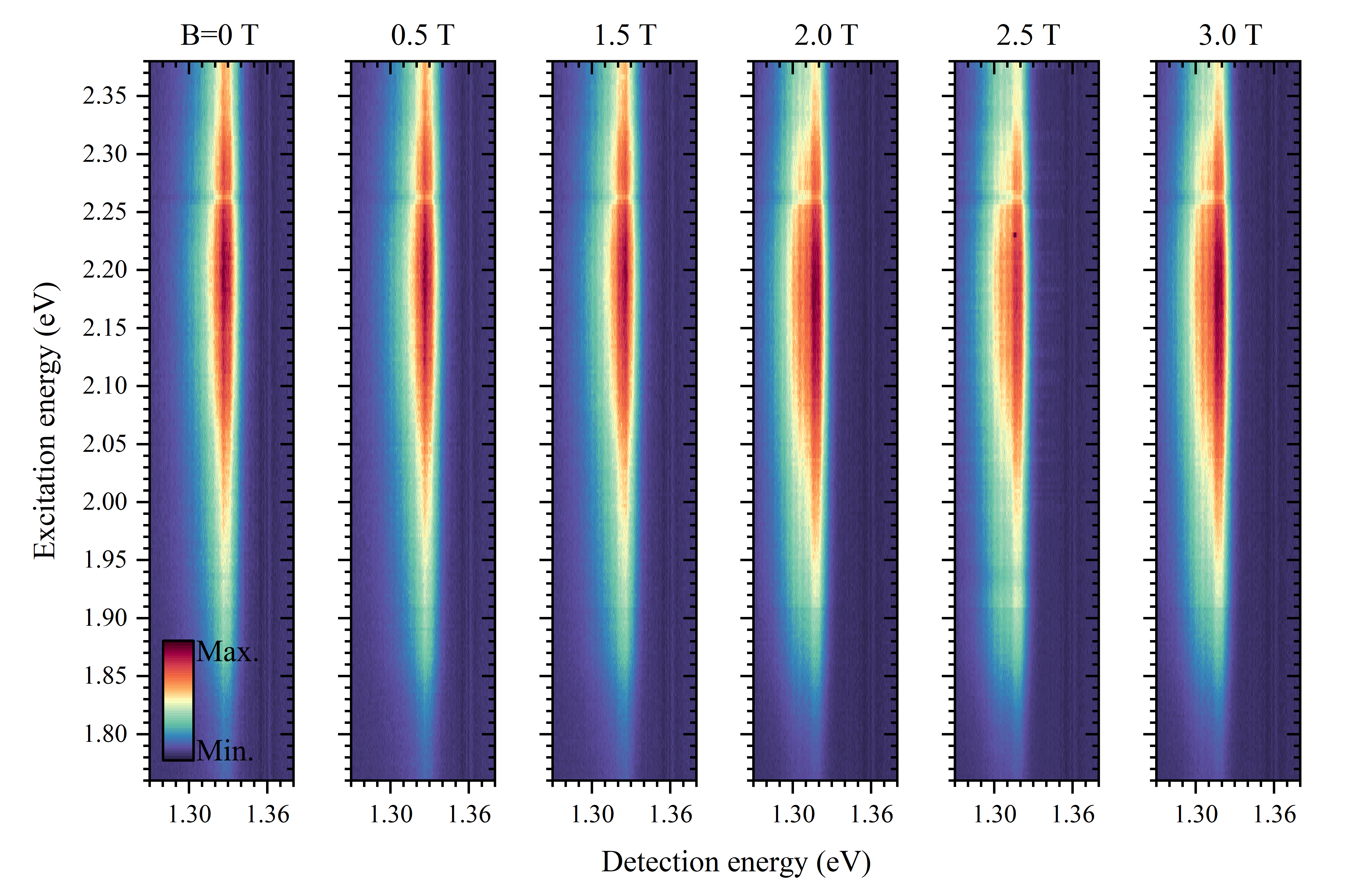}
    \caption{\label{si: figs. 6}
    Magnetic evolution PLE for 2L \crsbr.
    }
\end{figure}

The PLE maps at $T$=10~K of the AFM to FM transition in 1L (Fig.~\ref{si: figs. 5}) and 2L (Fig.~\ref{si: figs. 6}) \crsbr induced by out-of-plane $B_\textrm{ex}$.
This field orientation corresponds to the material's hard magnetization axis ($\hat{c}$).
Fig.~\ref{si: figs. 5} and ~\ref{si: figs. 5} show a set of PLE maps measured under different $B_\textrm{ex}$ (indicated on top of each panel) excited in the vicinity of the B and C transition, respectively.
The increase in the $B_\textrm{ex}$ leads to a redshift of both the A and A\textquotesingle.
However, in 1L sample peak intensity is significant affected.

\begin{figure}[h]
    \centering
    \includegraphics[width=.4\linewidth]{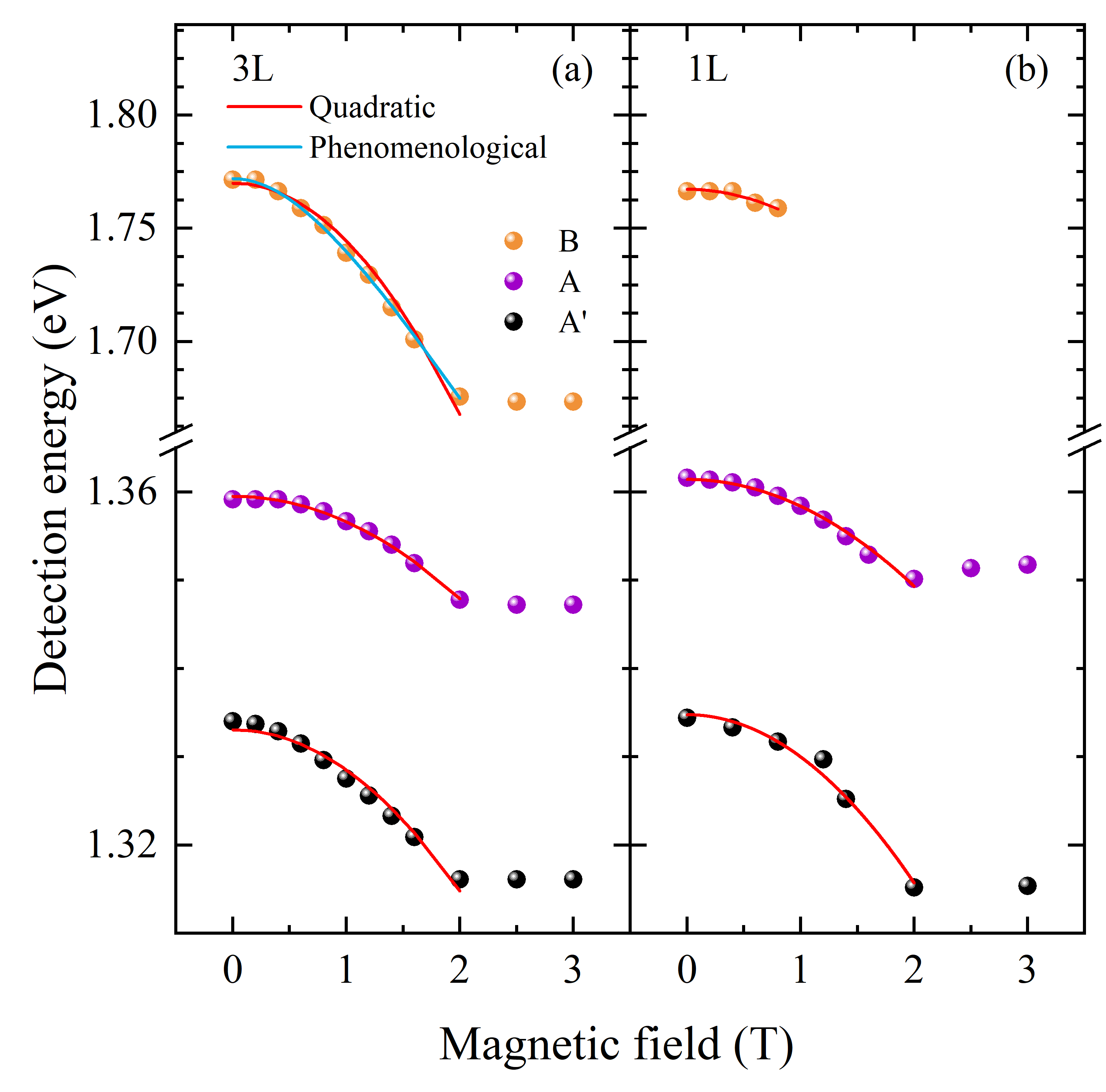}
    \caption{\label{si: figs. 7}
    Magnetic-dependent energy evolution for A, A\textquotesingle, and B for \crsbr~3L (a) and 1L (b).
    }
\end{figure}

The shift due to $B_\textrm{ex}$ of A, A\textquotesingle, and B for 3L (Fig.~\ref{si: figs. 7}(a)) and 1L (Fig.~\ref{si: figs. 7}(b)) \crsbr (2L is presented in main text) where described by the quadratic dependence of energy:
\begin{equation}
    E(B_\textrm{ex})=E_0-tB_\textrm{ex}^2
\end{equation}

\noindent where $E_0$ is the exciton energy and $t$ is related to both the interlayer hole tunneling and the energy splitting between the intralayer and interlayer excitons~\cite{tabataba2024, Christin2025}.
The fitted function is shown in Fig.~\ref{si: figs. 7} (red line) and the determined parameters are present in Table~\ref{Tbs: T1}.
However, the 3L B excitation is not precisely described.
To described such giant shift the approach proposed by Komar $et~al.$~\cite{Komar2024} was employed, where a phenomenological description of the states in terms of field-induced coupling yields:
\begin{equation}
    E(B_\textrm{ex})=\frac{1}{2}\left( E_1 \pm E_2-\sqrt{(E_1-E_2)^2+(\alpha B_\textrm{ex})^2} \right)-\beta B_\textrm{ex}^2
\end{equation}

\noindent where $E_2$ and $E_1$ are the zero-field energy of B and its high energy shoulder, respectively, $\alpha B_\textrm{ex}$ is the field-induced coupling, and $\beta B_\textrm{ex}$ originates from coupling with remote bands.
The fitted function was used to fit B exciton in 3L and 1L data, the results are shown in Fig.~\ref{si: figs. 7} (blue line) and the determined parameters are present in Table~\ref{Tbs: T2}.

\clearpage

\begin{table*}[]
\caption
    {
    Quadratic dependence of energy fittings parameters.
    }
\centering
\renewcommand{\arraystretch}{1.1}
\begin{tabular}[t]{ccc|cc|cc}
      & \multicolumn{6}{c}{$E(B_\textrm{ex})=E_0+tB_\textrm{ex}^2$}\\
      & \multicolumn{2}{|c|}{1L}& \multicolumn{2}{c|}{2L} & \multicolumn{2}{c}{3L} \\
      \hline
      & \multicolumn{1}{|c}{[eV]} & [eV/T$^2$] & [eV] & [eV/T$^2$] & [eV] & [eV/T$^2$] \\
      &  \multicolumn{1}{|c}{$E_0$} & $t$ & $E_0$ & $t$ & $E_0$ & $t$ \\
      \hline \hline
      A  &  \multicolumn{1}{|c}{1.361} & 0.0030 & - & - & 1.360 & 0.0029 \\
      A\textquotesingle &  \multicolumn{1}{|c}{1.335} & 0.0048 & 1.322 & 0.0033 & 1.333 & 0.0046 \\
      B &  \multicolumn{1}{|c}{1.767} & 0.013 & - & - & 1.770 & 0.0256 \\
      C &  \multicolumn{1}{|c}{-} & - & 2.198 & 0.0039 & - & - \\
      \end{tabular}
\label{Tbs: T1}
\end{table*}

\begin{table*}[]
\caption
    {
    phenomenological description of the states in terms of field-induced coupling~\cite{Komar2024} parameters.
    }
\centering
\renewcommand{\arraystretch}{1.1}
\begin{tabular}[t]{ccccc|cccc}
      & \multicolumn{8}{c}{$E(B_\textrm{ex})=\frac{1}{2}\left( E_1+E_2-\sqrt{(E_1-E_2)^2+(\alpha B_\textrm{ex})^2} \right)-\beta B_\textrm{ex}^2$}\\
      & \multicolumn{2}{|c|}{1L}& \multicolumn{2}{c|}{3L} \\
      \hline
      & \multicolumn{1}{|c}{[eV]} & [eV] & [eV/T] & [eV/T$^2$] & [eV] & [eV] & [eV/T] & [eV/T$^2$] \\
      &  \multicolumn{1}{|c}{$E_1$} & $E_2$ & $\alpha$ & $\beta$ & $E_1$ & $E_2$ & $\alpha$ & $\beta$ \\
      \hline \hline
      B &  \multicolumn{1}{|c}{1.82} & 1.177 & 0.038 & 0.086 & 1.93 & 1.77 & -0.157 & -0.016\\
      \end{tabular}
\label{Tbs: T2}
\end{table*}

\clearpage

\bibliographystyle{apsrev4-2}
\bibliography{biblio}